# Büttiker probes for dissipative phonon quantum transport in semiconductor nanostructures

K. Miao,[1,2] S. Sadasivam,[3] J. Charles,[1,2] G. Klimeck,[1,2] T. S. Fisher,[3] T. Kubis[2]

[1]School of Electrical and Computer Engineering, Purdue University, West Lafayette, Indiana 47907, USA
[2]Network for Computational Nanotechnology, Purdue University, West Lafayette, Indiana 47907, USA
[3]School of Mechanical Engineering and Birck Nanotechnology Center, Purdue University, West Lafayette, Indiana 47907, USA
Email: kmiao@purdue.edu

Theoretical prediction of phonon transport in modern semiconductor nanodevices requires atomic resolution of device features and quantum transport models covering coherent and incoherent effects. The nonequilibrium Green's function method (NEGF) is known to serve this purpose well but is numerically expensive in simulating incoherent scattering processes. This work extends the efficient Büttiker probe approach widely used in electron transport to phonons and considers salient implications of the method. Different scattering mechanisms such as impurity, boundary, and Umklapp scattering are included, and the method is shown to reproduce the experimental thermal conductivity of bulk Si and Ge over a wide temperature range. Temperature jumps at the lead/device interface are captured in the quasi-ballistic transport regime consistent with results from the Boltzmann transport equation. Results of this method in Si/Ge heterojunctions illustrate the impact of atomic relaxation on the thermal interface conductance and the importance of inelastic scattering to activate high-energy channels for phonon transport. The resultant phonon transport model is capable of predicting the thermal performance in the heterostructure efficiently.

State-of-the-art semiconductor logic and optoelectronic devices such as nanotransistors, quantum well photodetectors, and cascade lasers have characteristic device dimensions within the nanometer length scale [1, 2]. All of these devices can experience intense Joule heating and phonon generation during operation. Well established methods to describe phonon transport commonly employ the Boltzmann transport equation (BTE) [3, 4] or molecular dynamics [5, 6]. At the nanometer length scale, however, these methods exclude important quantum phenomena such as tunneling, confinement and interference effects [7-9]. For Fermionic nanoscale transport such as transport of charge and spin, the non-equilibrium Green's function method (NEGF) has proven to describe coherent and incoherent quantum effects accurately [10-12]. For phonon transport, the NEGF method has been predominantly used in the coherent regime [13-16], mainly due to the fact that inclusion of scattering typically requires self-consistent iterations of the NEGF equations. Such iterative solutions (e.g., the self-consistent Born approximation) are numerically demanding both in memory and CPU time, consequently restricting their application to impractically small nanostructures (Refs. [17] and [18] are notable exceptions). Büttiker probes [19,20,21] offer a heuristic but computationally efficient alternative to the self-consistent Born approximation; this model allows direct fitting to experimental transport data while avoiding self-consistent iterations within the NEGF equations [17].

In this work, the Büttiker probe approach that has been widely used for electronic transport is extended to model incoherent scattering in the atomistic phonon NEGF implementation of the nanodevice simulation tool NEMO5 [22]. The phonon Büttiker probe model is designed to mimic the frequency and temperature dependencies of phonon scattering rates and material-related thermal transport characteristics. As an example, this work shows results for a Ge/Si heterostructure. The physics underlying this model are illustrated and discussed for a generic material with varying effective scattering strength. The relevance of quantum transport with structural relaxation is also demonstrated through predictions of resonant phonon transport in a quasi-one dimensional Ge/Si heterostructure.

All silicon and germanium phonons in this work are represented by the modified valence force field model (MVFF) [23, 24]. The MVFF model includes the coupling energies of atoms with their first and second nearest neighbors in the material volume and third nearest neighbors in the plane of the first and second neighbors (i.e., third nearest neighbor co-planar coupling). All MVFF parameters are taken from Ref. [23]. Parameters at Si/Ge interfaces are arithmetical averages of the bulk Si and Ge parameters. Based on the potential energy $U$ and the harmonic approximation, the dynamical matrix ($D$) is calculated as

$$D_{mn}^{ij} = \frac{\partial^2 U}{\partial r_m^i \partial r_n^j}, i,j \in N_A \text{ and } m,n \in [x,y,z], \quad (1)$$

where $N_A$ is the number of atoms in the active device, $r_m^i$ and $r_n^j$ are the coordinates of atoms $i$ and $j$, and $x$-$y$-$z$ are the three spatial directions. If not stated otherwise, Si/Ge interfaces are relaxed within the MVFF model to minimize the potential energy $U$.

The nonequilibrium Green's function method requires the solution of a set of partial differential equations that for the case of phonons become [15, 17, 20]:

$$G^R(E) = \left[E^2 - D - \Sigma_{contact}^R - \Sigma_{BP}^R\right]^{-1}, \quad (2)$$

$$G^<(E) = G^R[\Sigma_{contact}^< + \Sigma_{BP}^<]G^{R\dagger}, \quad (3)$$

where $\Sigma_{contact}^R = \Sigma_{left}^R + \Sigma_{right}^R$ is the sum of the contact self-energies representing the device's connection to the left and right phonon reservoirs at the device/lead interface [10]. In stationary calculations of nanowires, Green's functions and self-energies depend on the phonon energy $E$ only. For periodic devices (e.g., ultrathin body, UTB, transistors) each periodic boundary condition perpendicular to the transport direction contributes an additional momentum term to the parameterization of Green's functions and self-energies. For a given parameter set (e.g., a given energy and momentum in a UTB transistor), the phonon Green's functions and self energies have three degrees of freedom for each atom. It is common to assume for the 'lesser' contact self-energy $\Sigma^<_{contact}$ that phonon occupancies in the leads are determined from equilibrium Bose distribution functions:

$$\Sigma^<_{contact} = \frac{1}{e^{E/k_B T_{left}}-1}\left(\Sigma^R_{left} - \Sigma^{R\dagger}_{left}\right) + \frac{1}{e^{E/k_B T_{right}}-1}\left(\Sigma^R_{right} - \Sigma^{R\dagger}_{right}\right) \quad (4)$$

where $T_{left}$ and $T_{right}$ are the temperatures of the left and right phonon reservoirs, respectively. The incoherent scattering of phonons is represented with one Büttiker probe self-energy per degree of freedom in the device:

$$\Sigma^R_{BP(i,m)} = -i\frac{2\omega}{\tau}, \quad (5)$$

$$\Sigma^<_{BP(i,m)} = \frac{1}{e^{E/k_B T_{(i.m)}}-1}\Sigma^R_{BP(i,m)}, \quad (6)$$

where $\tau$ is the scattering lifetime and $T_{(i.m)}$ is the temperature of the Büttiker probe at atom $i$ and polarization direction $m$. The retarded scattering self-energy is also included in the retarded contact self-energy calculation, following Refs. [17,25]. The retarded scattering self-energy is chosen to represent phonon-phonon Umklapp scattering $\tau^{-1}_{p-p} = C\omega^2 T^\alpha$, impurity scattering $\tau^{-1}_{IM} = B\omega^4$, and grain boundary scattering $\tau^{-1}_b = v_s/LF$ in the relaxation time approximation [4,26,27]. The total relaxation rate is assumed to be additive so that $\tau^{-1} = \tau^{-1}_b + \tau^{-1}_{IM} + \tau^{-1}_{p-p}$. The grain boundary scattering rate is determined from the average phonon group velocity $v_s$, the Casimir boundary length $L$, and a geometry factor , all taken from Ref. [4]. The scattering rates on impurities $\tau^{-1}_{IM}$ and on other phonons $\tau^{-1}_{p-p}$ depends on the constants $B$, $C$, and $\alpha$, respectively, which are chosen to best reproduce the thermal conductivity of silicon and germanium [4,28]. The temperature dependence of the phonon-phonon scattering is captured in this model (see Table I).

TABLE I. Fitting parameters for the Büttiker probes in Eq. (5)

| Material | $B[s^3]$ | $C[sK^{-\alpha}]$ | $\alpha$ |
|---|---|---|---|
| Si | $0.71\times10^{-45}$ | $1.74\times10^{-21}$ | 1.64 |
| Ge | $3.2\times10^{-45}$ | $1.12\times10^{-20}$ | 1.48 |

The temperature of each Büttiker probe $T_{(i.m)}$ is solved iteratively by Newton's method [22] until the integrated energy current

$$I_{(i,m)} = \int_{-\infty}^{+\infty} \sum_{j=1, j \neq i}^{N_A} \sum_{n=1, n \neq m}^{3} \frac{1}{h} E \left[ \mathcal{T}_{(i,m)(j,n)}(E) \left( \frac{1}{e^{E/k_B T_{i,m}}-1} - \frac{1}{e^{E/k_B T_{j,n}}-1} \right) + \mathcal{T}_{(i,m)left}(E) \left( \frac{1}{e^{E/k_B T_{i,m}}-1} - \frac{1}{e^{E/k_B T_{left}}-1} \right) + \mathcal{T}_{(i,m)right}(E) \left( \frac{1}{e^{E/k_B T_{i,m}}-1} - \frac{1}{e^{E/k_B T_{right}}-1} \right) \right] dE \quad (7)$$

vanishes for each Büttiker probe (*i,m*). Here, $\mathcal{T}_{(i,m)(j,n)}$, $\mathcal{T}_{(i,m)left}$ and $\mathcal{T}_{(i,m)right}$ are the transmission functions between Büttiker probes (*i,m*) and (*j,n*), and the left and right reservoir, respectively [20].

If not stated otherwise, all results in this work include phonon scattering on grain boundaries, impurities and other phonons. The scattering model of Eqs. (5-7) is similar to the electronic Büttiker probe model of Ref. [20] and implies that the limit for the phonon thermalization at each degree of freedom in the device (i.e., for each atom and each polarization direction) tends toward the respective local Büttiker probe temperature. The actual local temperature can differ from that of the Büttiker probe depending on the scattering strength, and we elaborate on this issue further in the next section. An estimate of the local temperature can be extracted from the average phonon gas energy:

$$\langle E \rangle = \int E^2 G^<(E) dE \approx \int E^2 \frac{A}{e^{E/k_B T_{local}}-1} dE. \quad (8)$$

where $A$ is the spectral function $A = i(G^R - G^{R\dagger})$. In this way, an effective, locally thermalized phonon gas of temperature $T_{local}$ is assumed to match the phonons solved in the NEGF equations with respect to their average energy. The average phonon mean free path $\lambda$ can be extracted following the approach of Ref. [17].

Figure 1 compares the experimental thermal conductivity of silicon and germanium of Refs. [4,28] with that computed by our NEGF model with Büttiker scattering of Eqs. (2-7). Table I shows the fitted scattering parameters that reproduce the experimental thermal conductivity over a large temperature range. The inset of Fig. 1 shows the average phonon mean free paths of silicon and germanium, which at 300K are 130nm and 40nm, respectively, and agree well with prior results [7, 30].

Figures 2 compares the Büttiker probe temperatures with the local temperature $T_{local}$ of Eq. (8). The Büttiker probe temperatures are averaged over the three polarization directions *m* as a function of atom position along the transport direction, $\sum_{m=1}^{3} T_{(i.m)}/3$. We note that the polarization dependence of the Büttiker probe temperature derives from scattering strengths and phonon dispersion. By construction, the Büttiker probe temperature is the target temperature of the phonon thermalization processes in the device. The results reveal that stronger scattering produces a larger temperature drop within the device and correspondingly smaller phonon conductance. The temperature jumps near the lead/device interfaces are more pronounced, with smaller scattering rates in the device region. Such jumps can be as attributed to intrinsic contact resistance. This resistance is not due to the scattering of phonons at the the lead/device interfaces since the contacts are reflectionless. Instead, the temperature jumps represent the non-equilibrium nature of ballistic transport. Alternative lead models that assume drifted distribution functions [25,32] give smaller temperature jumps at the lead/device interface. Similar discussion of the intrinsic contact resistance can be found in Ref. [34] where the temperature profile was obtained as a solution to Fourier's law and the Boltzmann transport equation. Accordingly, the temperature jumps at ideal contacts are captured by applying the proper boundary conditions to the heat diffusion equation.

Figure 3 shows the local temperature profile within several nanometers around a Si/Ge interface for various scattering strengths. Similar to Fig. 2, the left and right reservoir temperatures are 305K and 300K, and stronger scattering of the phonons in the bulk materials produces larger temperature gradients in the volume of the individual materials, with correspondingly smaller temperature jumps across the interfaces.

Si and Ge atoms must be relaxed near the interface due to their lattice mismatch in order to minimize the lattice energy. Figure 4 shows the importance of structural relaxation on the thermal interface conductance. The phonon thermal conductance of the relaxed structure is about 10% higher than in the unrelaxed case. As the scattering rate ratio between Si and Ge increases, the discrepancy between the relaxed and the unrelaxed structure reduces. Similar results occur for two-interface systems in which the thin Si layer is sandwiched between semi-infinite Ge leads.

Inelastic scattering is known to fill lower energy confined states in electron transport [1, 35]. This scattering-induced state-filling also holds for phonon transport as illustrated in Fig. 5 for a 3nm Ge layer embedded between two 3nm thick Si layers and Ge leads with 305K and 300K for the left and right lead temperatures. Figures 5(a) and 5(b) compare the energy-resolved phonon current and density of states of: (a) ballistic transport and (b) with transport that includes the Büttiker probes of Table I. In both cases, a large portion of phonon current is carried by phonons with energies similar to optical phonons of Ge (around 35meV). This result indicates that high energy phonons can coherently tunnel through the structure, and therefore the phonon current for these two cases are similar. The contour lines of the energy-resolved local phonon density of states in Fig. 5 show a higher density of states around 16meV in the central region when scattering is present. This density of states represents confined phonons of the central Ge layer that do not couple directly to the leads and therefore do not appear in the ballistic calculation. Inelastic scattering with phonons of different energies allows such confined modes to interact with the leads which (in this configuration) compensates the reduction of the net heat current caused by scattering in the other Ge areas (ballistic heat flux = $1.41 \times 10^{-9}$ W/nm$^2$; heat current with scattering $1.37 \times 10^{-9}$ W/nm$^2$). Similar findings of confined modes appearing with scattering in nanowire heterostructures were reported in Ref. [17].

In conclusion, a heuristic scattering quantum transport method of the NEGF with Büttiker probes is extended from the electronic framework to phonon transport. Relaxation time approximation models are used to describe the various scattering mechanisms in phonon transport. The scattering parameters of Büttiker probes in Si and Ge are fit to reproduce the experimental thermal conductivity of bulk Si and Ge over a large temperature range. Parametric studies have been performed on the effect of Büttiker probe scattering strengths on temperature jumps at interfaces as well as the deviation of local lattice temperature from the Büttiker probe temperature. The relaxation of atoms at and surrounding interfaces of different crystals is found to have a significant influence on the thermal conductance. The importance of inelastic scattering to enable the participation of higher energy confined states is illustrated with a Si/Ge heterostructure. The atomistic transport model introduced here enables the simulation of incoherent quantum phonon transport with a reasonable computational effort. The inelastic scattering changes the distribution of phonon current in the energy range.

Support under the Office of Naval Research (Award No. N000141211006, PM: Dr. Mark Spector) is gratefully acknowledged, as is support under SRC task 2141, SRC task 2273, by nanohub.org, and by the U.S. NSF (Nos. EEC- 0228390, OCI-0832623 and OCI-0749140).

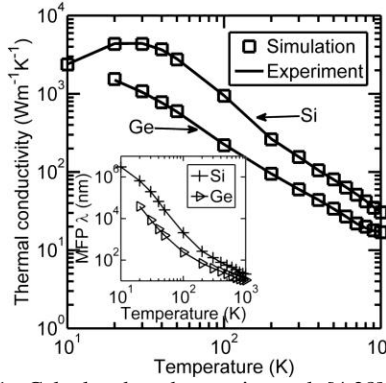

Figure 1. Calculated and experimental [4,28] thermal conductivity in the [100] direction as a function of temperature for Si and Ge. The Büttiker probe parameters in Table I are fitted to obtain agreement between the model and experimental data. The inset shows the corresponding mean free path.

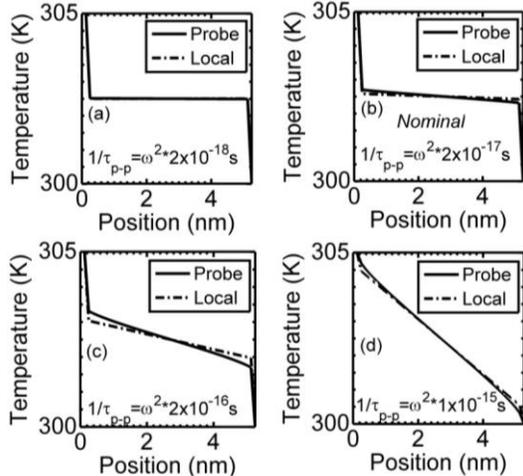

Figure 2. Calculated local and Büttiker probe temperature in a homogeneous 5.8 nm thick Si slab between Si leads of 305K and 300K when only phonon-phonon scattering is included and the scattering rate is varied. The transport direction is along the [100] crystal axis. Quasi ballistic (a) and nominally scattered (b) transport calculations show almost constant local temperature. Transport calculations with 10x (c) and 50x (d) larger than nominal scattering show pronounced temperature drops over the device.

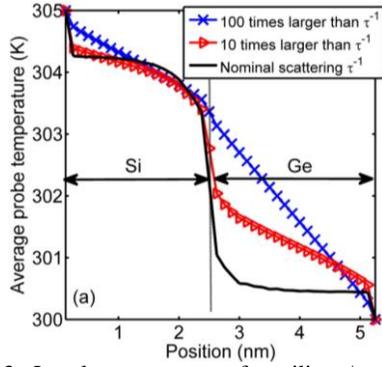

Figure 3. Local temperature of a silicon/germanium interface connected to homogeneous semi-infinite Si and Ge leads of 305K and 300K phonon temperature, respectively. The transport direction is along [100] crystal direction. The scattering rates of Si and Ge are varied with constant factors (10x triangles, and 100x crosses) in Si and Ge simultaneously.

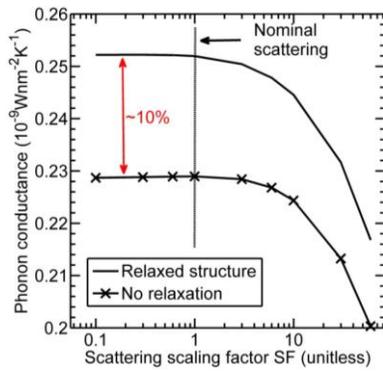

Figure 4. Phonon conductance of the Si/Ge interface of Fig. 3 with nominal scattering potentials scaled with a factor SF for both Si and Ge simultaneously. SF equals the ratio of the actual scattering rate to the fitted scattering rate. The thermal conductance of the strain-relaxed interface exceeds the conductance when all atoms are fixed on a Ge native lattice.

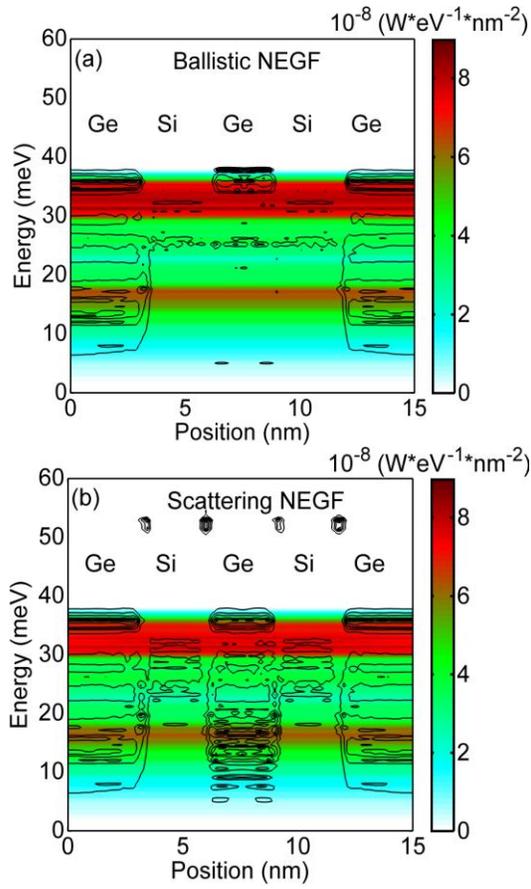

Figure 5. Contour plots of the energy and position resolved phonon energy current densities of a 3nm Ge layer separated by two 3nm Si layers from semi-infinite homogeneous Ge grown in [100] direction. The contour lines represent the density of states. The right and left lead temperatures are 305K and 300K, respectively. The ballistic situation (a) lacks a pronounced density of states of confined phonon modes in the central Ge layer that a calculation including scattering parameters of Table I shows (b).